\documentclass[aps,prl,twocolumn,groupedaddress,showpacs,floatfix]{revtex4}
\usepackage[utf8]{inputenc}
\usepackage[portuges]{babel}
\usepackage{dcolumn}
\usepackage{amsmath}
\usepackage{amssymb}
\usepackage{graphicx}
\usepackage{bm}
\usepackage[T1]{fontenc}
\usepackage{color}
\usepackage{url}
\usepackage[bf]{subfigure}
\usepackage{rotating}
\usepackage{color}

\begin{document}

\title{Explosive synchronization enhanced by time-delayed coupling}

\author{Thomas Kau\^{e} Dal'Maso Peron}
\affiliation{Instituto de F\'{\i}sica
de S\~{a}o Carlos, Universidade de S\~{a}o Paulo, Av. Trabalhador
S\~{a}o Carlense 400, Caixa Postal 369, CEP 13560-970, S\~{a}o
Carlos, S\~ao Paulo, Brazil}
\author{Francisco A. Rodrigues}
\email{francisco@icmc.usp.br}
\affiliation{Departamento de Matem\'{a}tica Aplicada e Estat\'{i}stica, Instituto de Ci\^{e}ncias Matem\'{a}ticas e de Computa\c{c}\~{a}o,
Universidade de S\~{a}o Paulo - Campus de S\~{a}o Carlos, Caixa Postal 668,
13560-970 S\~{a}o Carlos, SP, Brazil.}

\begin{abstract}
This paper deals with the emergence of synchronization in scale-free networks by considering the Kuramoto model of coupled phase oscillators. The natural frequencies of oscillators are assumed to be correlated with their degrees and a time delay is included in the system. This assumption allows enhancing the explosive transition to reach a synchronous state. We provide an analytical treatment developed in a star graph which reproduces results obtained in scale-free networks. Our findings have important implications in understanding the synchronization of complex networks, since the time delay is present in most systems due to the finite speed of the signal transmission over a distance.
\end{abstract}



\maketitle

\section{Introduction}

Synchronization is an emerging collective phenomenon present in science, nature, social life and engineering~\cite{Pikovsky03, Strogatz03, Nadis03:Nature}. When a large population of limit-cycle oscillators is coupled, with slight different natural frequencies, a synchronous state emerges and those oscillators run at an identical frequency~\cite{Pikovsky03, Arenas08:PR}. For instance, experimental studies have verified that neurons in the central nervous system synchronize their oscillatory responses~\cite{Gray89:Nature}. Similarly, the emergence of synchronization has also been observed among electrochemical oscillators~\cite{Kiss02:Science}. Indeed, the onset of synchronization has been verified in circadian rhythm, communication networks, power grids, social interactions, cortical networks and ecology~\cite{Arenas08:PR}.

Winfree proposed a mathematically tractable model to explain the emergence of collective synchronization~\cite{Winfree67}. Although his approach allows describing the synchronization dynamics in a population of coupled oscillators, it assumes that every oscillator feels the same mean-field, which is hard to be observed in real-world complex systems. A more suitable model was proposed by Kuramoto, who considered oscillators coupled by the sine of their phase differences and phase oscillators at arbitrary frequencies~\cite{Acebron05:RMP, Moreno04:EPL}. Therefore, each oscillator $i$ obeys an equation of motion given by
\begin{equation}
\frac{d\theta_{i}}{dt}= \omega_i + \lambda \sum_{i=1}^{N}A_{ij} \sin(\theta_j - \theta_i), \quad i=1,\ldots,N,
\end{equation}
where $\lambda$ is the coupling strength, $\omega_i$ is the natural frequency of oscillator $i$ (generally distributed according to some function $g(\omega)$), and $A_{ij}$ are the elements of the adjacency matrix $A$, which represents the topology of a complex network. More specifically, elements $A_{ij} = 1$ if two nodes $i$ and $j$ are connected, and $A_{ij} = 0$, otherwise~\cite{Costa011:AP}. In the Kuramoto model, each node tries to oscillate independently at its own frequency, while its direct neighbors tend to synchronize it to all the other oscillators. It has been verified that coupling strengths higher than a determined threshold $\lambda_c$ produce the onset of synchronization~\cite{Arenas08:PR}. The level of synchronization of the whole system is measured by the macroscopic complex order parameter~\cite{Arenas08:PR},
\begin{equation}
r(t) e^{i \psi(t)} = \frac{1}{N} \sum _{j=1}^{N} e^{i \theta_{j}(t)}.
\label{eq:coherence}
\end{equation}
where $0 \leq r(t) \leq 1$ measures the phase coherence of populations. When $r(t)=1$, the system reaches the complete synchronous state and when $r(t)=0$ it reaches an asynchronous state.

The Kuramoto model has also been analyzed considering some modifications, such as different types of coupling functions and frequency distributions~\cite{Motter05:EPL, Arenas08:PR}, in order to enhance the synchronization level. For instance, Garde\~{n}es et. al.~\cite{Gardenes011:PRL} investigated the Kuramoto model in complex networks by defining the natural frequencies of each node $i$ as degree $k_{i}$, \emph{i.e.}, $\omega_{i}=k_{i}$. In contrast to the canonical choice for the frequency distribution, they observed a first-order phase transition to the synchronous state in scale-free networks, showing that this effect is due to the positive correlation between the nodes frequencies and the network topology.

The Kuramoto model has also been modified to include time-delay (\emph{e.g.}~\cite{Yeung99:PRL, Choi00:PRE, Peres011:Chaos, Peres011:PRE}).
Time delay is observed in communication systems due to the finite speed of the signal transmission over a distance. For instance, the transmission of signals through unmyelinated axons presents an 80-ms delay for propagation through a cortical network~\cite{Kandel}. As verified by Dhamala \emph{et al.}~\cite{Dhamala04:PRL}, the time delay can enhance the network synchronization. They studied neural synchrony by taking into account the Hindmarsh-Rose neurons and observed that an extended region of stable synchronous activity could be achieved with low coupling strengths only if the delay were included in the system. The contribution of the time delay to network synchronization was also verified  by P\'{e}rez \emph{et al.} ~\cite{Peres011:Chaos}. The authors incorporated it in the Kuramoto model and verified that it was possible to obtain a perfect synchronization in directed networks by relating the topology of the network and the phases and frequencies of the oscillators.

In the current paper, we took into account the analyse by Garde\~{n}es et. al.~\cite{Gardenes011:PRL} and Choi \emph{et al.}~\cite{Choi00:PRE} altogether. We considered the natural frequencies of each node $i$ as the node degree $\omega_{i}=k_{i}$ and included a time delay in the Kuramoto model. Therefore, the set of equations of motion for $N$ coupled oscillators is given by
\begin{equation}
\frac{d\theta_{i}}{dt} = \omega_{i} + \lambda \sum_{j=1}^{N} A_{ij} \sin\left[ \theta_{j}(t-\tau) - \theta_{i}(t) \right].
\label{eq:kuramoto_delay}
\end{equation}
Note that the phase of node $i$  interact with the phase of node $j$ with a time delay $\tau$.

\begin{figure}[t]
\centerline{\includegraphics[width=1\linewidth]{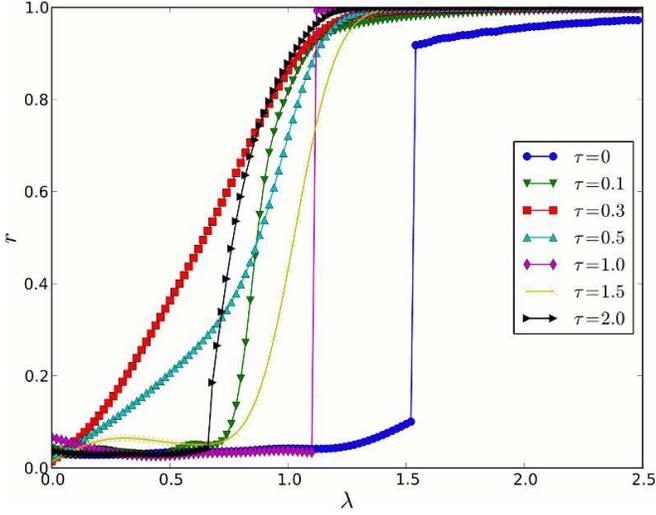}}
\caption{Coherence diagram $r(\lambda)$ for BA model with $N=1000$ and $\langle k \rangle = 6$.}\label{Fig:BA_delay}
\end{figure}

We analyzed this modified Kuramoto model (Eq.~\ref{eq:kuramoto_delay}) in Barab\'{a}si-Albert (BA) networks~\cite{Barabasi99:Science}. Fig.~\ref{Fig:BA_delay} shows the synchronization diagram $r(\lambda)$ taking into account a scale-free topology with $N=10^{3}$ and $\langle k \rangle=6$. As performed in~\cite{Gardenes011:PRL} we increased the value of $\lambda$ adiabatically and computed the stationary value of the global coherence $r$ for each value $\lambda_{0},\; \lambda_{0} + \delta \lambda,\; \ldots ,\; \lambda_{0}+ n\delta \lambda $. Fig.~\ref{Fig:BA_delay} shows that the presence of a time delay in the phase interactions allows reaching the synchronous state with smaller couplings than without time delay. In addition, it is interesting to observe that the system presents different phase transitions for each time-delay $\tau$ adopted. For instance, for $\tau = 1.0$, the transition is of first order, while for $\tau = 0.3$, $r$ increases almost linearly with $\lambda$ and no phase transition is verified.

In order to locally investigate the effect of the time delay on BA networks, a star graph was considered as an approximation of the scale-free topology. A star graph is a special type of tree, \emph{i.e.}, a central node connected with $K$ leaves. Thus, for a star network with $N=K+1$ nodes, the peripheral node has degree $k_{i}=1 \; (i=1,..., K)$ and the central node has $k_{H} = K$ connections. By using this approximation we studied the influence of hubs (nodes with higher frequencies) on neighbors nodes, which are expected to have lower frequencies, since the average clustering coefficient tends to zero in BA networks~\cite{Boccaletti06:PR, Costa07:AP}.
As in the work by Garde\~{n}es et. al.~\cite{Gardenes011:PRL}, we considered the natural frequencies of each node equal to its respective degree. Setting the rotated frame with the average phase $\psi(t) = \psi(0) + \Omega t$, where $\Omega$ is the average frequency of the system, we defined the new variables as $\phi_{i} = \theta_{i} - \Omega t$ for the peripheral nodes and $\phi_{H} = \theta_{H} - \Omega t$ for the
hub. These definitions allowed rewriting Eq.~\ref{eq:kuramoto_delay} into the set of equations given by
\begin{equation}
\dot{\phi}_{H} = (\omega_{H} - \Omega) + \lambda \sum_{j=1}^{K} \sin\left[ \phi_{j}(t-\tau) - \phi_{H}(t) - \Omega \tau \right],
\label{eq:star_hub}
\end{equation}
\begin{equation}
\dot{\phi}_{i} = (\omega - \Omega) + \lambda \sin\left[ \phi_{H}(t-\tau) - \phi_{i}(t) - \Omega \tau \right],
\label{eq:star_leave}
\end{equation}
where we set $\psi(0)=0$ without loss of generality. The global parameter defined in Eq.~\ref{eq:coherence} can be rewritten into
\begin{equation}
r = \frac{e^{i \phi_{H}} + \sum_{j=1}^{K} e^{i \phi_{j}} }{K+1}.
\label{eq:coherence_2}
\end{equation}
Multiplying both sides of Eq.~\ref{eq:coherence_2} by $e^{-i(\phi_{H} + \Omega \tau)}$ and taking the imaginary part, we can rewrite Eq.~\ref{eq:star_hub} as
\begin{equation}
\dot{\phi}_{H} = (\omega_{H} - \Omega) - \lambda r(K+1)\sin(\phi_{H} + \Omega \tau) - \lambda \sin\Omega \tau .
\label{eq:hub_mean_field}
\end{equation}
Imposing the phase locked solution $\dot{\phi}_{H}=0$ we obtained
\begin{equation}
\sin(\phi_{H}+\Omega \tau ) = \frac{(\omega_{H} - \Omega )}{\lambda r (K+1)} + \frac{\sin{\Omega \tau}}{r(K+1)}.
\end{equation}\label{eq:sin_H}
Note that when the time delay vanishes, $\tau =0$, the same result verified in~\cite{Gardenes011:PRL} for the hub's phase is recovered (\emph{i.e.}\ $\sin \phi_{H} = (\omega_{H}-\Omega)/\lambda r (K+1)$). Considering Eq.~\ref{eq:star_leave} and solving it for $\cos \phi_{i}$ in the locked regime $\dot{\phi_{i}}=0$ we obtained
\begin{equation}
\begin{split}
\cos{\phi_{i}} = \frac{ (\Omega - \omega)\sin(\phi_{H} - \Omega \tau)}{\lambda} \pm \\\frac{\sqrt{ \left[1 - \sin^{2}(\phi_{H} -\Omega \tau) \right]\left[ \lambda^{2} - (\omega - \Omega)^{2} \right] } }{\lambda}.
\end{split}
\label{eq:cos_i}
\end{equation}

This equation is valid only for $\lambda \geqslant \left|\omega - \Omega\right|$, which leads to the critical coupling $\lambda_{c} = \left|\omega - \Omega\right|$ for the onset of synchronization. Nevertheless it is important to stress that this value for the critical coupling depends on the time delay $\tau$, once the average frequency also depends on it (\emph{i.e.}, $\Omega = \Omega(\tau)$). When $\tau=0$, the result $\Omega = 2K/(K+1)$ verified in~\cite{Gardenes011:PRL} is recovered. Moreover, it is straightforward to show that at the critical coupling $\lambda = \lambda_{c}$ the phases of the peripheral nodes and the hub obey the following condition
\begin{equation}
\delta \phi(\tau) = \phi_{H} - \phi_{i} = \frac{\pi}{2} + \tau \Omega(\tau).
\label{eq:phases_diff}
\end{equation}
The above equation shows that the difference between the phases of peripheral nodes depends directly on the time delay $\tau$, which means that it is possible to tune the time delay in order to obtain different phase couplings between the peripheral nodes and the hub.

\begin{figure}[!t]
\begin{center}
\centerline{\includegraphics[width=1.0\linewidth]{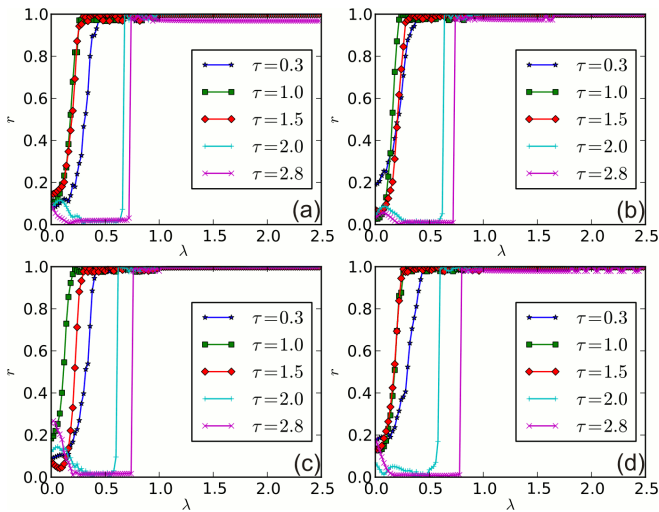}}
\caption{Coherence diagram $r(\lambda)$ for (a)$K=60$, (b)$K=70$, (c)$K=80$ and (d)$K=90$ }
\label{Fig:Star_delays}
\end{center}
\end{figure}

In order to obtain insights on how the presence of the time delay $\tau$ is related to the onset of synchronization, we constructed the synchronization diagram for the star networks, as shown in Fig.~\ref{Fig:Star_delays}. Comparing the results presented in Figures~\ref{Fig:Star_delays} and \ref{Fig:all_K_inplot}(inset), we can see that the time-delay enhance the network synchronization. The values of $K$ were selected by taking into account the expected degree for a hub in a scale-free network of $N=10^{3}$, as we considered the star graph an approximation of a BA network. Remarkably, Fig.~\ref{Fig:Star_delays} shows that star graphs present the same behavior observed in the scale-free network in the presence of a time delay (see Fig.\ref{Fig:BA_delay}). More specifically, depending on the value of $\tau$, the star networks undergo different synchronization transitions. It is interesting to note that, like in Yeung et. al.~\cite{Yeung99:PRL}, a subcritical Hopf bifurcation of the incoherent state at $\tau=1$ is observed. In addition, a supercritical Hopf bifurcation at $\tau=2$ is verified in Figure~\ref{Fig:BA_delay}, for the BA networks.

\begin{figure}[!t]
\begin{center}
\centerline{\includegraphics[width=1\linewidth]{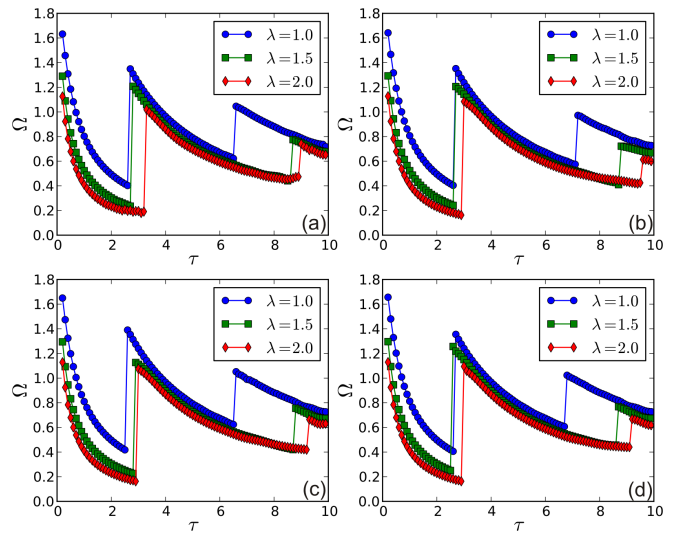}}
\caption{Average frequency $\Omega$ as a function of time delay $\tau$ for (a)$K=60$, (b)$K=70$, (c)$K=80$ and (d)$K=90$ }
\label{Fig:OmegaXtau}
\end{center}
\end{figure}

This fact is due to the dependence of the mean frequency $\Omega$ on $\tau$, \emph{i.e.}, $\Omega = \Omega(\tau)$. We analyzed such dependence in order to understand how the critical coupling $\lambda_{c}$ varies for different time delays. Fig.~\ref{Fig:OmegaXtau} shows the mean frequency $\Omega$ as a function of $\tau$.
It is interesting to note that the particular choice of the frequency distribution $g(\omega)=P(k)$ leads to a dependence of $\Omega(\tau)$ that is different than that verified in~\cite{Choi00:PRE}, in which the authors found that frequency $\Omega$ depends inversely on the time delay $\tau$ for the Kuramoto model with all-to-all connectivity. Fig.~\ref{Fig:OmegaXtau} shows an oscillatory dependence between $\Omega$ and $\tau$, which explains the different behavior of the synchronization diagrams displayed by the scale-free and stars networks in contrast to the case without time delay~\cite{Gardenes011:PRL} (see Figs.~\ref{Fig:BA_delay} and~\ref{Fig:Star_delays}). In these figures, the value of $\lambda$ to reach the onset of synchronization varies periodically with $\tau$.
For instance, in Fig.~\ref{Fig:BA_delay}, $\lambda_c(\tau = 1.0)<\lambda_c(\tau = 0.3)<\lambda_c(\tau = 1.5)$, while in Fig.~\ref{Fig:Star_delays}, $\lambda_c(\tau = 1.0) <\lambda_c(\tau = 0.3)<\lambda_c(\tau = 2.0)$.

\begin{figure}[!t]
\centerline{\includegraphics[width=1.0\linewidth]{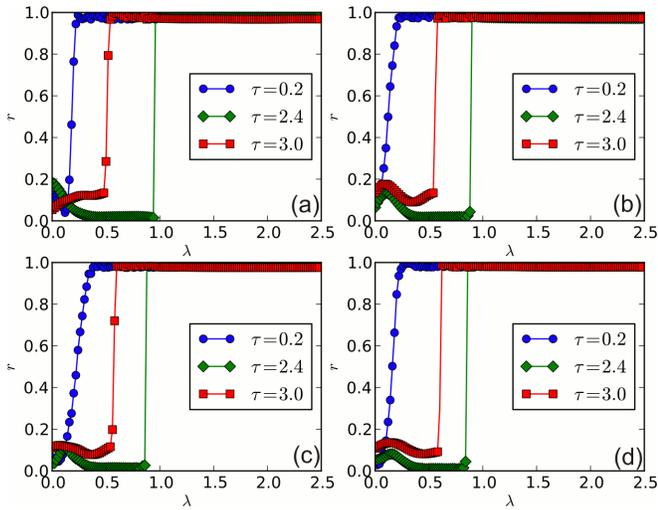}}
\caption{Coherence diagram $r(\lambda)$ for (a)$K=60$, (b)$K=70$, (c)$K=80$, (d)$K=90$.}
\label{Fig:stars_0_pi}
\end{figure}

Given this oscillatory dependence between $\Omega$ and $\tau$, we determine the value of $\tau$ that makes the star networks more susceptible to the emergence of the onset of synchronization. More specifically, we found a time delay value that enhances the network synchronization level, while keeping the state stable in the thermodynamic limit $K\rightarrow\infty$. We also determined a value of time delay that yields the maximum value of critical coupling, \emph{i.e.}, a time delay that decreases the synchronization level. From equation Eq.~\ref{eq:cos_i}, we obtained $\lambda_c = |\omega - \Omega|$. Therefore, while higher frequencies lead to a lower critical coupling enhancing the network synchronization, lower frequencies lead to higher ones, disturbing the network synchronization. In order to exam those extreme cases, Figure~\ref{Fig:stars_0_pi} shows the coherence diagram for time delays chosen accordingly to Figure~\ref{Fig:OmegaXtau}. For instance, for the time delay $\tau=2.4$, the system presents the lowest possible frequency, in such a way that the network undergoes a first-order transition to the synchronous state. Nevertheless, for the time delay $\tau=0.2$, no phase transition is observed and the order parameter $r$ is increased until the synchronous state has been reached.

Note that if the time delay $\tau$ is set in order to establish a higher critical coupling $\lambda_{c}$, a first-order phase transition is observed. Also, differently to the case in which $\tau=0$~\cite{Gardenes011:PRL}, the dependence of the critical coupling $\lambda_{c}$ with the number of nodes is lost. More specifically, in~\cite{Gardenes011:PRL} the authors show that the stability of the unlocked state ($r \approx 0 $) increases as $K \rightarrow \infty $, while here we observed that the transition to the synchronous state does not depend extensively on $K$ (see Fig.~\ref{Fig:all_K_inplot}).

\begin{figure}[!tpb]
\centerline{\includegraphics[width=0.8\linewidth]{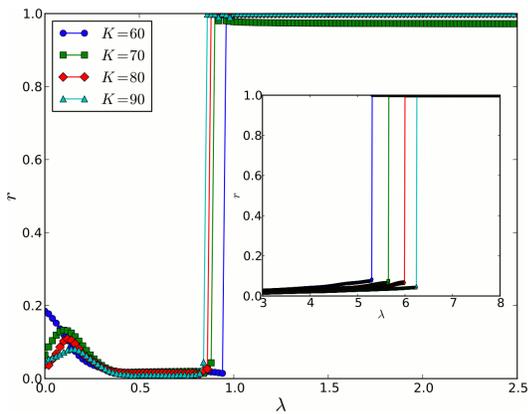}}
\caption{Comparison of the explosive synchronization observed in star networks with different sizes $K$. The same delay shown in Figure~\ref{Fig:OmegaXtau} was adopted for the worst cases. The inset shows the same stars networks but for the case with $\tau=0$.}
\label{Fig:all_K_inplot}
\end{figure}

\begin{figure}[!t]
\centerline{\includegraphics[width=1.0\linewidth]{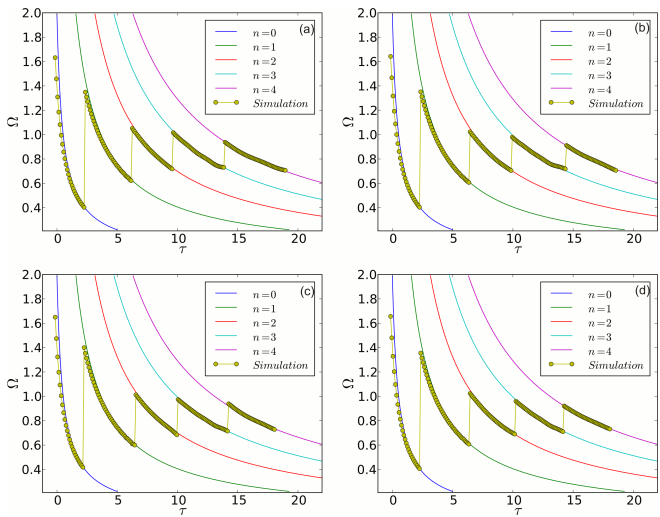}}
\caption{Theoretical and numerical curves for the frequency $\Omega$ as function of $\tau$ with $\lambda=1.0$, for the cases: (a)$K=60$, (a)$K=70$, (c)$K=80$ and (d)$K=90$. The discrete points correspond to the data shown in Figure~\ref{Fig:OmegaXtau}.}
\label{Fig:freq_curves}
\end{figure}

\begin{figure}[!t]
\centerline{\includegraphics[width=1.0\linewidth]{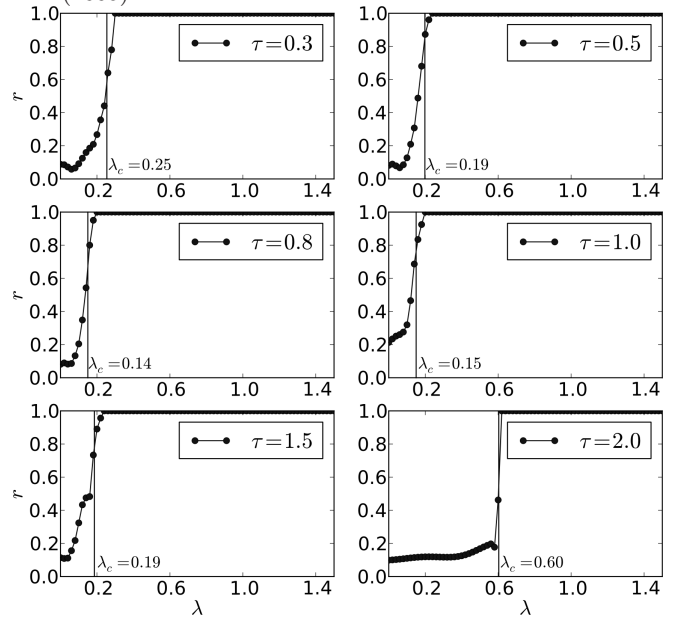}}
\caption{Synchronization diagrams for a star network with $K=80$ leaves and theoretical estimation for the critical coupling $\lambda_{c}$ for each time delay displayed.}
\label{Fig:critical_coupling}
\end{figure}

Let us now investigate more deeply the frequency dependency  $\Omega=\Omega(\tau)$ supposing
\begin{equation}
\phi_{hub,l}(t) = \theta_{hub,l} - \Omega t = \alpha_{hub,l},
\label{eq:ansatz}
\end{equation}
where we have used the stationary solution with a phase difference $\theta_{hub,l}=\Omega_{hub,l} t + \alpha_{hub,l}$. From Eq.~\ref{eq:ansatz}, we can write Eq.~\ref{eq:star_hub} and~\ref{eq:star_leave} for the $K+1$ nodes in the star network and thus leading to the following system of $K+1$ equations
\begin{eqnarray*}
\Omega & = & K+\lambda\sum_{j=1}^{K}\sin(-\Omega\tau+\alpha_{i} - \alpha_{H})\\
\Omega & = & 1-\lambda\sin(\Omega\tau+\alpha_{1}-\alpha_{H})\\
\Omega & = & 1-\lambda\sin(\Omega\tau+\alpha_{2}-\alpha_{H})\\
 & \vdots\\
\Omega & = & 1-\lambda\sin(\Omega\tau+\alpha_{K}-\alpha_{H}).
\label{eq:sytem_eq}
\end{eqnarray*}
The average frequency $\Omega$ of the star networks must satisfy all these equations. Summing all the equations in the system we have
\begin{equation}
(K+1)\Omega=2K+\lambda\sum_{j=1}^{K}\left[\sin(-\Omega\tau+\alpha_{i})-\sin(\Omega\tau+\alpha_{i})\right].
\label{after_system}
\end{equation}
Setting $\alpha_{H}=\alpha_{1}=\alpha_{2}=\cdots=\alpha_{N}$ without loss of generality we obtain
\begin{equation}
\Omega = \frac{2K}{K+1}\left[ 1 - \lambda\sin(\Omega \tau) \right].
\label{eq:trans_Omega}
\end{equation}
According to~\cite{earl2003synchronization} the system is linearly stable if and only if
\begin{equation}
\cos(\Omega \tau) > 0
\label{eq:cos_Otau}
\end{equation}
which means that the product $\Omega \tau$ must be in the range
\begin{equation}
2n\pi < \Omega\tau <(2n+1)\pi,\; n=0,1,2,...
\label{eq:stability_condition}
\end{equation}
Considering $\Omega \tau \rightarrow \Omega \tau - n\pi$ in Eq.~\ref{eq:trans_Omega}, in Figure~\ref{Fig:freq_curves} we show the solutions of Eq.~\ref{eq:trans_Omega} for different values of $n$. Observe that these solution correspond to the average frequency shown in Figure~\ref{Fig:OmegaXtau}. Therefore, we verified a high agreement between theoretical and simulated results. Furthermore, once we know exactly how the average frequency $\Omega$ depends on $\tau$, is possible to determine the critical coupling for the star networks through the expression $\lambda_{c} = \left|\omega - \Omega \right|$ and with Eq.~\ref{eq:stability_condition} by rewriting Eq.~\ref{eq:trans_Omega} just in terms of $\lambda_{c}$ and $\tau$. Fig.~\ref{Fig:critical_coupling} displays the results obtained for the critical couplings $\lambda_{c}$ for different values of $\tau$ for the star network with $K=80$, showing a perfect agreement for the theoretical values of $\lambda_{c}$ with the numerical data.

In conclusion, the addition of time delay to the Kuramoto model allows reaching a higher synchronization level than those observed by Garde\~nes \emph{et al.}~\cite{Gardenes011:PRL}. Even for a small time delay, the onset of synchronization is obtained for smaller coupling strengths than those for networks without time delay. The analysis performed here helps to understand real world communication systems, in which the time delay is present due to the finite speed of the signal transmission across a communication medium.

Francisco A. Rodrigues would like to acknowledge CNPq (305940/2010-4) and FAPESP (2010/19440-2) for the financial support given to this research. Thomas K. D. M. Peron would like to acknowledge Fapesp for the sponsorship provided. The authors also acknowledge Angela C. P. Giampedro, who provided a careful review of the text.

\bibliographystyle{apsrev}
\bibliography{paper}

\end{document}